\documentclass[prl,aps,reprint,superscriptaddress]{revtex4-2}
\usepackage{bbm}
\usepackage{amsmath,amssymb}
\usepackage{physics}
\usepackage{dsfont}
\usepackage{graphicx}
\usepackage{hyperref}
\usepackage[dvipsnames]{xcolor}
\usepackage{bm}

\begin{document}

%Title of paper
\title{Is the real two-Higgs-doublet model consistent? }
\def\Carleton{Ottawa-Carleton Institute for Physics, Carleton University, Ottawa, ON K1S 5B6, Canada}
\def\TRIUMF{TRIUMF, 4004 Wesbrook Mall, Vancouver, BC V6T 2A3, Canada}

\author{Carlos Henrique de Lima}
\email{cdelima@triumf.ca}
\affiliation{\TRIUMF}
\affiliation{\Carleton}

\author{Heather E.\ Logan}
\email{logan@physics.carleton.ca}
\affiliation{\Carleton}

\date{\today}
% change
\begin{abstract}
We provide strong evidence that the widely-studied real two-Higgs-doublet model is inconsistent under renormalization due to quark-induced divergent CP violation (CPV). We identify the necessary ingredients for the CPV to enter the renormalization-group equations based on symmetry proprieties of the divergent diagrams. We demonstrate that while these ingredients are present starting at six loops, the divergent CPV is zero at that order due to approximate symmetries. We show that these symmetries are broken at seven loops and determine the parameter dependence of the resulting divergent CPV.
\end{abstract}

%\maketitle must follow title, authors, abstract, and keywords
\maketitle

%%%%%%%%%%%%%%%%%%%%%%%%%%%%%%%%%%%%%%%%%%%
\section{Introduction}
\label{sec:intro}
%%%%%%%%%%%%%%%%%%%%%%%%%%%%%%%%%%%%%%%%%%%

While the $125$~GeV Higgs boson discovered at the CERN Large Hadron Collider~\cite{higgsdiscovery1,higgsdiscovery2} consists mostly of a single $SU(2)_{L}$ doublet~\cite{ATLAS:2022vkf,CMS:2022dwd} as predicted by the Standard Model (SM), there is no fundamental reason to expect only one Higgs doublet. The most common extension compatible with data that appears in many well-motivated frameworks beyond the SM, including supersymmetric models~\cite{Dimopoulos:1981zb}, is the two-Higgs-doublet model (2HDM)~\cite{Lee:1973iz,Branco:2011iw}.

The 2HDM was originally introduced~\cite{Lee:1973iz} as a spontaneous source of CP violation (CPV), which generically causes asymmetries between the properties of particles and their antiparticles and intrinsic electric dipole moments of electrons and nuclei. While all observed CPV phenomena to date are well-described by the explicit CPV in the Cabibbo-Kobayashi-Maskawa (CKM) quark-mixing matrix~\cite{Kobayashi:1973fv}, removing the immediate need for a CPV 2HDM, new sources of CPV are needed to explain the baryon asymmetry of the universe~\cite{Sakharov:1967dj}.

The most popular version of the 2HDM imposes an additional softly-broken $Z_2$ symmetry to avoid flavor-changing neutral Higgs interactions~\cite{Glashow:1976nt,Paschos:1976ay}, giving rise to the familiar type-I and -II 2HDMs~\cite{Branco:2011iw}. While these models admit explicit CPV in the scalar potential~\cite{Gunion:2005ja}, most studies have imposed CP conservation in the scalar sector, yielding the model known as the real 2HDM.
 
It was recently pointed out by Fontes et al.~\cite{Fontes:2021znm} that the real 2HDM is expected to suffer from theoretical inconsistencies under renormalization. The CPV in the CKM matrix should radiatively induce CPV in the scalar potential. Since the CPV in the CKM matrix arises from the dimensionless Yukawa couplings, this radiative CPV should be divergent. Renormalizing the theory would then require the presence of counterterms with imaginary parts, which are not present in the original real 2HDM potential. If this indeed occurs, excluding the complex phases in the tree-level scalar potential would render the 2HDM non-renormalizable.

Reference~\cite{Fontes:2021znm} took the first step in understanding this problem by computing the leading divergence of the three-loop pseudoscalar $A^{0}$ tadpole in the type-II 2HDM. Unexpectedly, after summing the result over quark flavors, the coefficient of the leading divergence was found to be exactly zero. Our objective is to explain this cancelation and determine at what loop order a nonzero divergent CPV term should be expected to appear.

We approach this problem using a novel application of approximate symmetries to obtain results beyond the current reach of perturbative calculations. Working in the unbroken electroweak phase, we construct the minimal pair of primitive diagrams involving a closed quark loop, which, once the extra external scalar legs are closed, generate every possible diagram that could contain CPV from the quark sector at the lowest nontrivial order. This dramatically simplifies the analysis and highlights the important roles played by natural flavor conservation and other approximate symmetries of the model.

We find that all necessary ingredients for divergent CPV are present starting at the six-loop order. However, we identify additional approximate symmetries of the diagrams that lead to an exact cancelation of the divergent CPV piece. This cancelation happens for different reasons in the type-I and -II models. Extending the analysis to seven loops, we show that additional coupling insertions break all the identified approximate symmetries and that the cancelation ceases to occur. We thus expect divergent CPV contributions to the scalar potential at seven-loop order in both the type-I and -II 2HDMs, with different parametric dependence in the two models. Full details are provided in a companion paper~\cite{long}.

This analysis serves as a cautionary tale for model-building beyond the SM. It is a general feature of quantum field theory that if a sector of a model violates a symmetry, this violation will eventually propagate to the rest of the model. However, in most models, it is common to rely on analyzing the first few loop orders and assume that if the symmetry-breaking effects have not yet appeared, they will be safe to ignore. This work reinforces that such an assumption is generally false. 
This is not necessarily a phenomenological disaster, as the symmetry-breaking coefficient often remains small even if it is not technically natural~\cite{tHooft:1979rat}. 
However, it implies the need for a high-energy explanation for such small coefficients and complicates the analysis of such theories beyond the tree level.

%%%%%%%%%%%%%%%%%%%%%%%%%%%%%%%%%%%%%%%%%%%
\section{Necessary ingredients for CPV}
\label{sec:2HDM}
%%%%%%%%%%%%%%%%%%%%%%%%%%%%%%%%%%%%%%%%%%%

The quark Yukawa Lagrangian in the type-II 2HDM takes the form~\footnote{The lepton sector does not affect our analysis, trivially extending our results to types X and Y. In the companion paper~\cite{long}, we also study the model without natural flavor conservation, in which the CP leaks can occur at one-loop because of the flavor-changing neutral Higgs interactions.}
\begin{equation}
	\mathcal{L}_{Y} \, = \,  - (Y_d)_{ij} \overline{Q}_{Li} \Phi_1 d_{Rj} 
		- (Y_u)_{ij} \overline{Q}_{Li} \widetilde \Phi_2 u_{Rj} + {\rm h.c.},
	\label{eq:LYuk}
\end{equation}
where h.c.\ stands for Hermitian conjugate and $\widetilde \Phi \equiv \varepsilon \Phi^*$ with $\varepsilon$ being the two-index antisymmetric tensor.  Type I is obtained by replacing $\Phi_1$ with $\Phi_2$ in the first term. This Yukawa structure is enforced by imposing a $Z_2$ symmetry $\Phi_1 \to -\Phi_1$ with the right-handed quarks transforming appropriately~\cite{Glashow:1976nt,Paschos:1976ay}.

Applying the same $Z_2$ symmetry and allowing soft (dimension-two) breaking terms, the scalar potential is~\cite{Wu:1994ja,Branco:2011iw},
\begin{eqnarray}
V_H
&&
\,=\, 
m_{11}^2 \Phi_1^\dagger \Phi_1 + m_{22}^2 \Phi_2^\dagger \Phi_2
- \left[ m_{12}^2 \Phi_1^\dagger \Phi_2 + \mathrm{h.c.} \right]
\nonumber\\[2pt]
&&
+\, \tfrac{1}{2} \lambda_1 (\Phi_1^\dagger\Phi_1)^2
+ \tfrac{1}{2} \lambda_2 (\Phi_2^\dagger\Phi_2)^2
+ \lambda_3 (\Phi_1^\dagger\Phi_1) (\Phi_2^\dagger\Phi_2)
\nonumber\\[2pt]
&&
+\, \lambda_4 (\Phi_1^\dagger\Phi_2) (\Phi_2^\dagger\Phi_1)
+ \left[
\tfrac{1}{2} \lambda_5 (\Phi_1^\dagger\Phi_2)^2
+ \mathrm{h.c.}
\right]
  . \, \, \,  
\label{eq:scalarpot}
\end{eqnarray}
Here $m_{12}^2$ and $\lambda_5$ are in general complex parameters~\cite{Lavoura:1994fv,Botella:1994cs}.  Imposing an \emph{exact} $Z_2$ forces $m_{12}^2 = 0$; then the phase of $\lambda_5$ can be rotated away, and CP is conserved in the scalar potential. Once $m_{12}^2$ is allowed in the theory, the scalar potential generically contains CPV. The real 2HDM is defined by choosing $m_{12}^2$ and $\lambda_5$ to be real in the same basis. 

The Yukawa Lagrangian in Eq.~(\ref{eq:LYuk}) is also, by accident~\cite{Ferreira:2010ir}, invariant under a global $U(1)$ symmetry,
\begin{equation}
	\Phi_1 \to e^{-i \theta} \Phi_1, \qquad \Phi_2 \to e^{i \theta} \Phi_2,
\end{equation}
with the quarks transforming according to
\begin{eqnarray}
	&&u_R \to e^{i \theta} u_R, \qquad d_R \to e^{-i \theta} d_R \qquad {\rm (type~I)} \nonumber \\
	&&u_R \to e^{i \theta} u_R, \qquad d_R \to e^{i \theta} d_R \qquad \ \ {\rm (type~II)},
\end{eqnarray}
and $Q_L$ invariant.  This is equivalent to the Peccei-Quinn symmetry~\cite{Peccei:1977hh} for the type II case, so we refer to it as $U(1)_{PQ}$.

This $U(1)_{PQ}$ can be extended to Eq.~\eqref{eq:scalarpot}, where it forces $m_{12}^2 = \lambda_5 = 0$~\cite{Ferreira:2009wh}.  Under this exact $U(1)_{PQ}$, the scalar potential is CP conserving; including the soft breaking term $m_{12}^{2}$ does not change this fact because the phase of $m_{12}^2$ can be rotated away~\footnote{An explicit demonstration of this at one loop is given in Ref.~\cite{Pilaftsis:1998pe}.}. The coupling $\lambda_5$ thus acts as a spurion that breaks the would-be $U(1)_{PQ}$ down to $Z_2$. This gives us our first nontrivial piece of information: any diagram that would give rise to divergent CPV in the scalar sector must involve a $\lambda_{5}$ insertion.

All CPV observables arising from the CKM matrix $V$ can be parameterized in terms of a single quantity known as the Jarlskog invariant~\cite{Jarlskog:1985ht,Jarlskog:1985cw}, given by $J = \left| {\rm Im} ( V_{\alpha i} V_{\beta j} V^*_{\alpha j} V^*_{\beta i} ) \right|$ with $\alpha \neq \beta$, $i \neq j$. It is convenient to express this invariant in terms of the quark Yukawa matrices~\cite{Botella:1994cs,Silva:2004gz}. Defining $\widehat{H}_u =  Y_u Y_{u}^{\dagger}$ and $\widehat{H}_d =  Y_d Y_{d}^{\dagger}$, the minimal combination that yields an imaginary part (proportional to the Jarlskog invariant) is
\begin{align} \label{def:complex}
\mathcal{J}= \Tr \left( \widehat{H}_u  \widehat{H}_d  \widehat{H}_u^2  \widehat{H}_d^2 \right) \, ,
\end{align}
where the trace sums over the three generations. This quantity involves a total of 12 powers of Yukawa matrices.

\begin{figure}[h!]
 \resizebox{0.45\linewidth}{!}{ \includegraphics{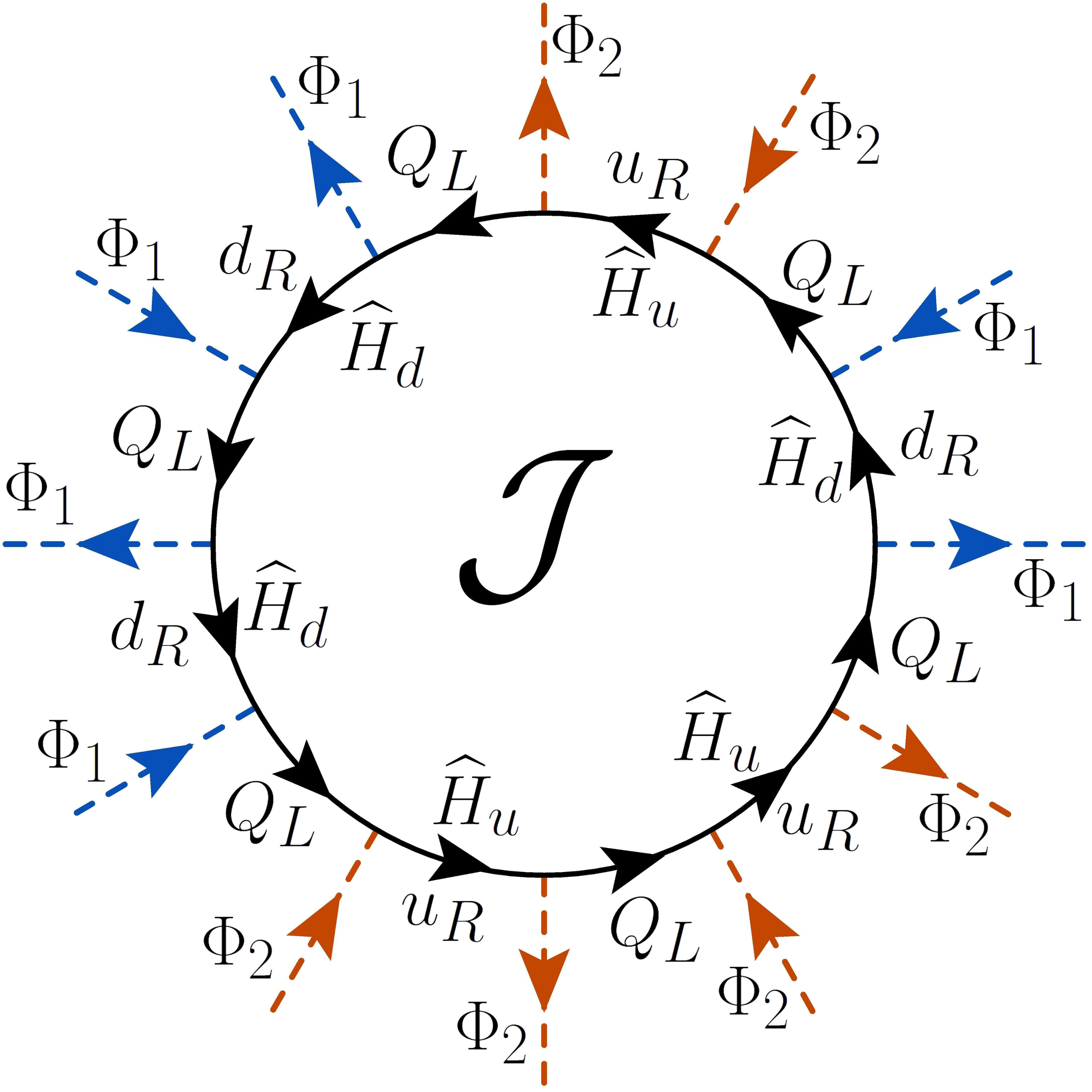}} \,  \resizebox{0.45\linewidth}{!}{ \includegraphics{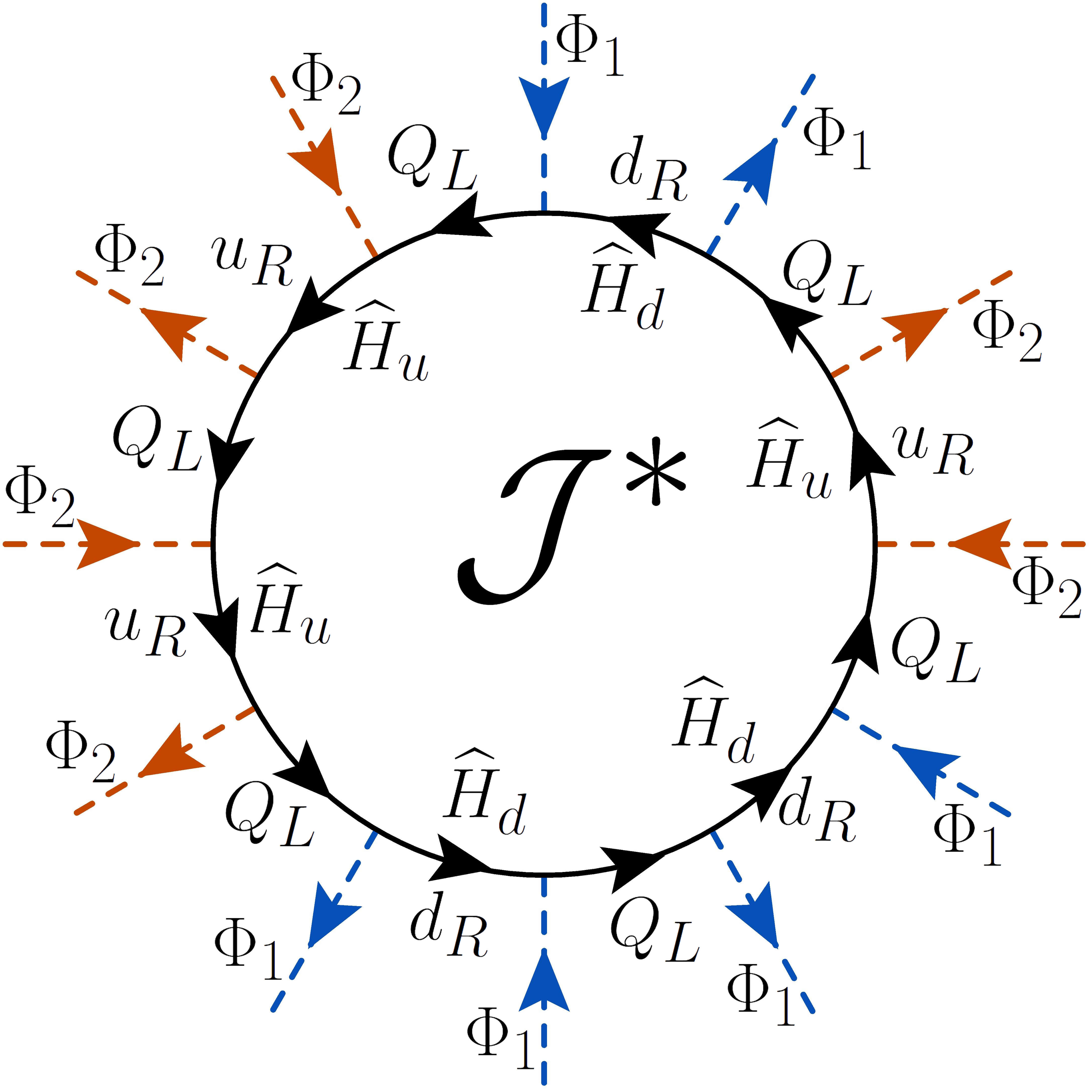}}
\caption{The smallest closed quark loops that yield the Jarlskog invariant via $\mathcal{J}$ and $\mathcal{J}^*$, shown for the type II 2HDM.  For type I, replace each $\Phi_1$ with $\Phi_2$. The $\mathcal{J}^*$ diagram is built from the $\mathcal{J}$ diagram by replacing $u_R \leftrightarrow d_R$ and adjusting the attached scalar lines accordingly; an alternative construction involves reversing the flow of fermion number.}
\label{fig:bigcircle}
\end{figure}

The form in Eq.~(\ref{def:complex}) allows us to construct the minimal pair of primitive diagrams proportional to $\mathcal{J}$ and $\mathcal{J}^*$, respectively, as shown in Fig.~\ref{fig:bigcircle}. This provides our second nontrivial piece of information: any diagram that would give rise to scalar-sector CPV originating from the CKM matrix must involve a quark loop proportional to $\mathcal{J}$ or $\mathcal{J}^*$. Together with the requirement of a $\lambda_5$ insertion, this implies that divergent CPV contributions to scalar potential parameters cannot appear below the six-loop level: connecting eight of the 12 scalar legs of either primitive diagram to produce a four-scalar operator creates a five-loop diagram, and inserting a $\lambda_5$ vertex requires one additional loop.

%%%%%%%%%%%%%%%%%%%%%%%%%%%%%%%%%%%%%%%%%%%
\section{Six- and seven-loop analysis}
\label{sec:noleak}
%%%%%%%%%%%%%%%%%%%%%%%%%%%%%%%%%%%%%%%%%%%

In order to show that the real 2HDM is theoretically inconsistent, we need to demonstrate the existence of a nonzero divergent contribution to $\Im(\lambda_{5})$. Starting from the two primitive diagrams of Fig.~\ref{fig:bigcircle} together with a $\lambda_5$ vertex, we can construct all diagrams that contribute to the $(\Phi_1^{\dagger} \Phi_2)^2$ operator by choosing all relevant configurations of external legs and connecting the remaining legs in all possible ways.

Calculating the divergences of generic six-loop diagrams is beyond the reach of current computational tools. We instead look for pairings of diagrams proportional to $\mathcal{J}$ and $\mathcal{J}^*$ that give rise to identical divergent contributions so that the imaginary part cancels in their sum. These pairings are driven by two transformations that convert the primitive diagram proportional to $\mathcal{J}$ into the one proportional to $\mathcal{J}^*$: these are ($i$) in the type-I model, reversal of the fermion flow, which forces $\Phi_2 \leftrightarrow \Phi_2^*$; and ($ii$) in the type-II model, interchange of $u_R \leftrightarrow d_R$, which forces $\Phi_1 \leftrightarrow \widetilde \Phi_2$. The first is the standard CP transformation applied to the entire primitive diagram; the second corresponds to a generalized CP transformation~\cite{Branco:2011iw}.  

The key observation is that because the coefficient of the four-scalar operator is dimensionless, the divergent parts of the diagrams cannot depend on any particle masses or external momenta. The transformations that take $\mathcal{J} \leftrightarrow \mathcal{J}^*$ are thereby promoted to accidental symmetries of the divergent parts of the Feynman integrals at six loops, so that the sum of each such pair of diagrams is proportional to $\left(\mathcal{J}+\mathcal{J}^{*}\right)$, and hence purely real. Since the additional accidental symmetry differs in types I and II, we study the two models separately.

%%%%%%%%%%%%%%%%%%%%%%%%%%%%%%%%%%%%%%%
\subsection{Type-I 2HDM}
%%%%%%%%%%%%%%%%%%%%%%%%%%%%%%%%%%%%%%%

\begin{figure}[h!]
 \resizebox{0.65\linewidth}{!}{ \includegraphics{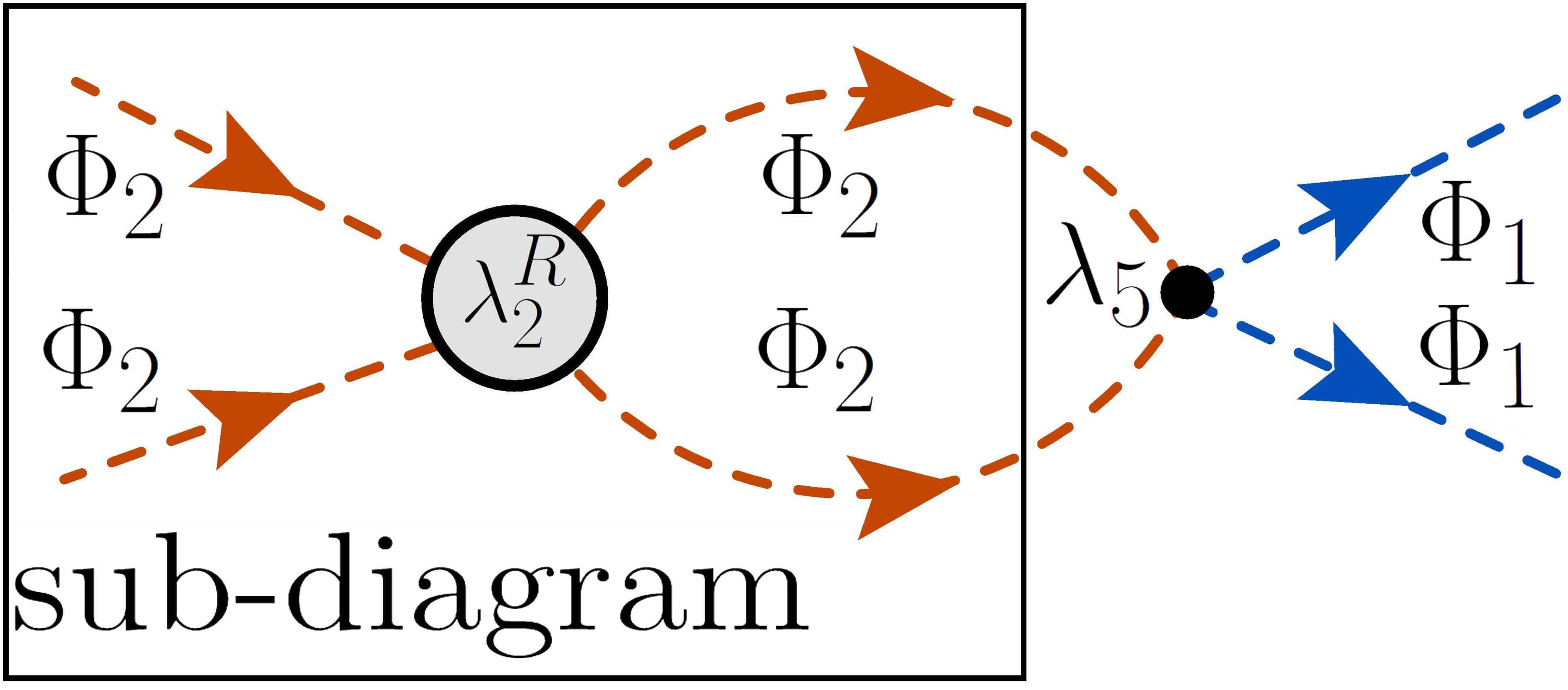}} \,
\caption{Sub-diagram topology of the six-loop diagrams in the type-I 2HDM.}
\label{fig:RG}
\end{figure}

In the type-I model, only $\Phi_2$ is connected to the fermion line. Since we want to generate the operator $(\Phi_1^{\dagger} \Phi_2)^2$, we must convert two outgoing $\Phi_{2}$ fields into $\Phi_{1}$ using the $\lambda_{5}$ insertion. All of the six-loop diagrams then have the characteristic topology shown in Fig.~\ref{fig:RG}, in which it is always possible to cut the two $\Phi_2$ propagators connected to the $\lambda_{5}$ vertex and thereby isolate a five-loop sub-diagram.

This gives us a clearer picture of the six-loop correction, as it is possible to integrate and sum all the sub-diagrams first and generate a five-loop form factor before performing the sixth loop integral. The symmetry transformation $\Phi_2 \leftrightarrow \Phi_2^*$, $q \to q^*$ (with $q = Q_L, u_R, d_R$) that interchanges the diagrams proportional to $\mathcal{J}$ and $\mathcal{J}^{*}$ also interchanges the incoming and outgoing $\Phi_2$ lines of the sub-diagram.  This forces the $\lambda_5$ insertion to be attached instead to the other pair of legs, changing the momentum structure of the overall diagram. 

The form factor from the five-loop sub-diagram can, in principle, depend nontrivially on which legs carry the momentum of the sixth loop. However, since any four-scalar form factor is dimensionless and the theory is effectively massless for the purpose of computing only the divergent parts, only the piece of the five-loop form factor that reduces to a constant in the zero-momentum limit can contribute. This constant piece must be real because it constitutes the five-loop renormalization of the coefficient $\lambda_2$ of the Hermitian operator $(\Phi_2^{\dagger} \Phi_2)^2$.  This guarantees that the sum of all six-loop diagrams proportional to $\mathcal{J}$ generates the same divergent coefficient as the sum of all such diagrams proportional to $\mathcal{J}^*$, so that the imaginary part of $\mathcal{J}$ cancels in their sum. \emph{Finite} imaginary contributions do not necessarily cancel, but they do not affect the renormalizability of the real 2HDM.
We thus conclude that the type-I 2HDM has no divergent CPV at six loops.

\begin{figure}[h!]
 \resizebox{0.485\linewidth}{!}{ \includegraphics{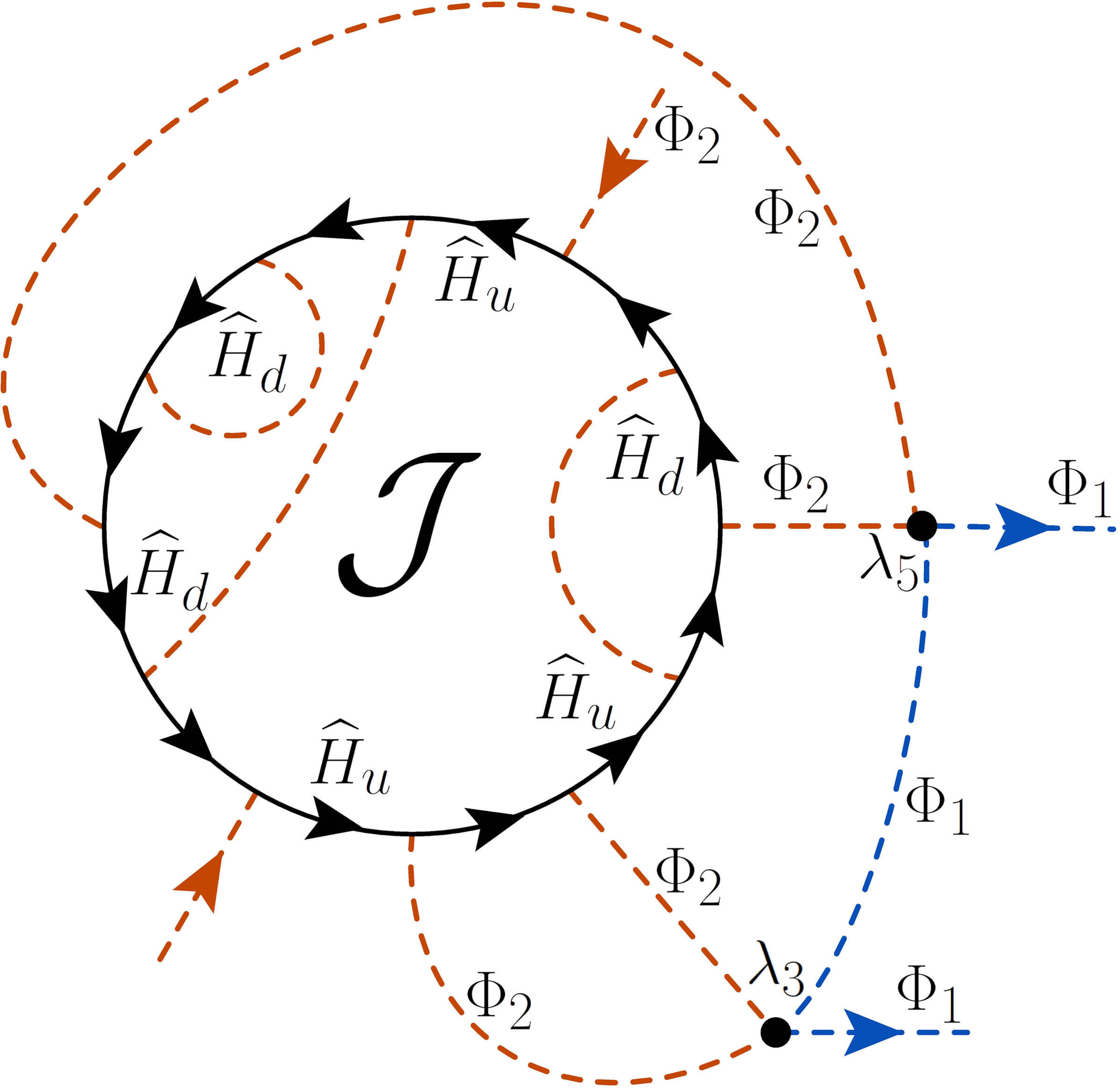}} \,  \resizebox{0.485\linewidth}{!}{ \includegraphics{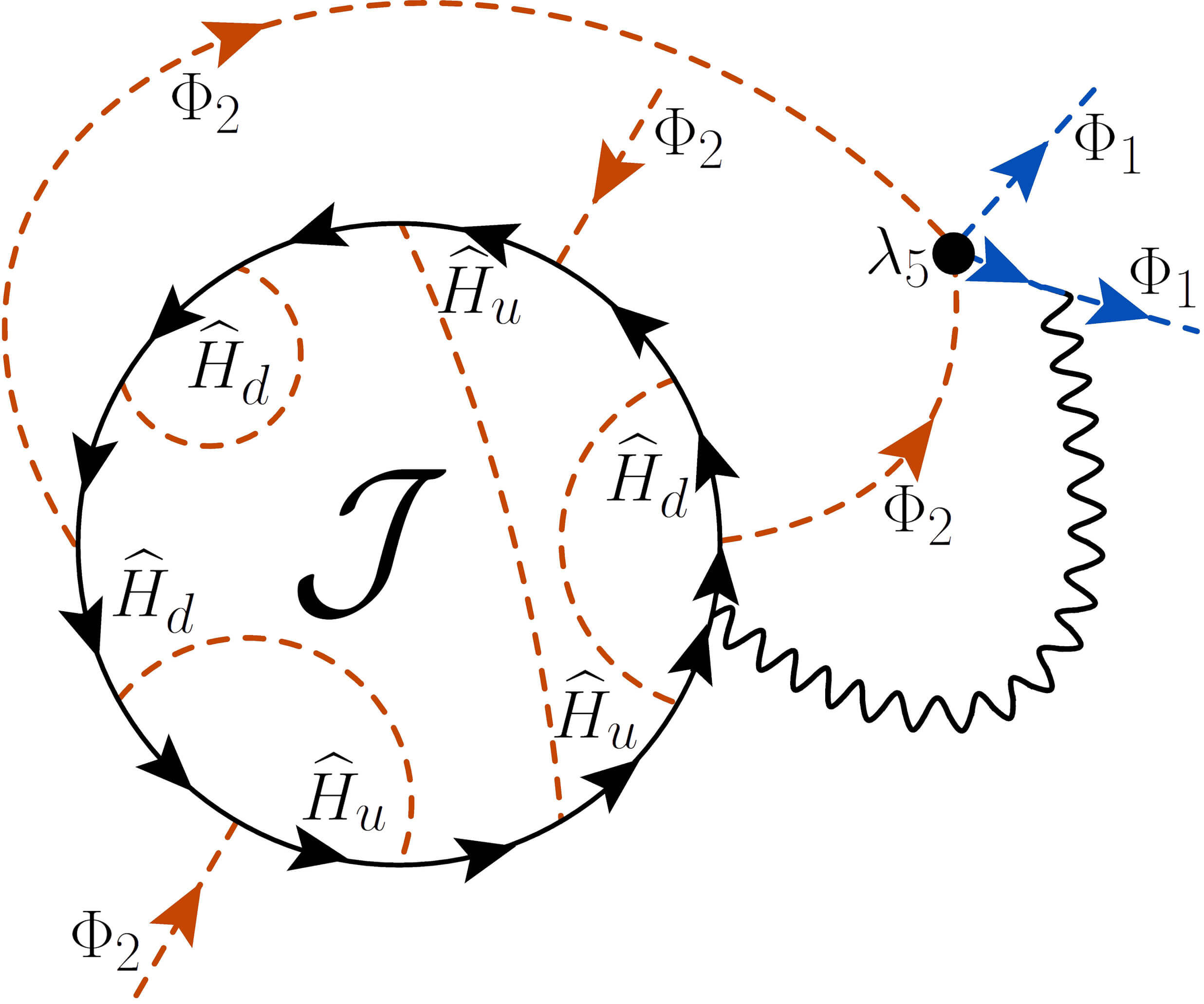}}
\caption{Two seven-loop diagrams in the type-I 2HDM that violate the sub-diagram topology.}
\label{fig:typeI7loop}
\end{figure}

This sub-diagram structure is not preserved at seven loops, and the accidental enhanced symmetry that ensures the cancelation of the divergent imaginary parts is thus removed. Diagrams without the sub-diagram structure can be constructed by connecting a propagator between the original sub-diagram and one of the external $\Phi_1$ legs, as shown in Fig.~\ref{fig:typeI7loop}. All such diagrams involve one of the following:
\begin{itemize}
\item a $\lambda_{3}$ or $\lambda_{4}$ vertex;
\item a hypercharge or $SU(2)_L$ gauge boson exchange.
\end{itemize}
We thus expect imaginary divergent contributions to $\lambda_5$ to first appear at the seven-loop level in the type-I 2HDM, with coefficients proportional to $\lambda_5 \Im(\mathcal{J})$ times these additional couplings. We have explicitly verified the breaking of the sub-diagram structure by generating the diagram topologies involving $\lambda_{3}$ or $\lambda_{4}$ using QGRAF~\cite{QGRAF}.

We finally note that diagrams contributing radiative corrections to $m_{12}^{2}$ are also guaranteed to take the form of a four-scalar sub-diagram with two legs connected to an $m_{12}^2$ insertion, and this holds to all orders and in both the type-I and -II models. Such corrections can, therefore, contain an imaginary divergent piece only if the sub-diagram is not Hermitian; i.e., the sub-diagram must already correspond to the operator $(\Phi_1^{\dagger} \Phi_2)^2$. This shows that the leading imaginary divergent contribution to $m_{12}^2$ always occurs at one higher loop order than the leading imaginary divergent contribution to $\lambda_5$, thereby ensuring that the phase of $\lambda_5$ thus generated cannot be trivially rotated away~\cite{Gunion:2005ja}.

%%%%%%%%%%%%%%%%%%%%%%%%%
\subsection{Type-II 2HDM}

In the type-II model, every diagram proportional to $\mathcal{J}$ that contributes to the four-scalar operator $(\Phi_1^{\dagger} \Phi_2)^2$ can be converted into a diagram proportional to $\mathcal{J}^*$ by performing the generalized CP transformation,
\begin{align}\label{eq:GEN}
\Phi_1 \leftrightarrow \widetilde{\Phi}_2
\,  \, , \, \, u_R \leftrightarrow d_R \, .
\end{align} 
This transformation preserves the four-scalar operator $(\Phi_1^{\dagger} \Phi_2)^2$ in the limit of zero external momenta. It affects the $SU(2)_L$ index structure by interchanging $\varepsilon \leftrightarrow \varepsilon^{T}$ in the product of Yukawa vertices; however, since there are the same number of $\varepsilon$ and $\varepsilon^{T}$ insertions, this does not generate a relative minus sign between the diagrams. This immediately demonstrates that the imaginary divergent contribution to $\lambda_5$ at six loops cancels between pairs of diagrams related by this transformation.

\begin{figure}[h!]
 \resizebox{0.375\linewidth}{!}{ \includegraphics{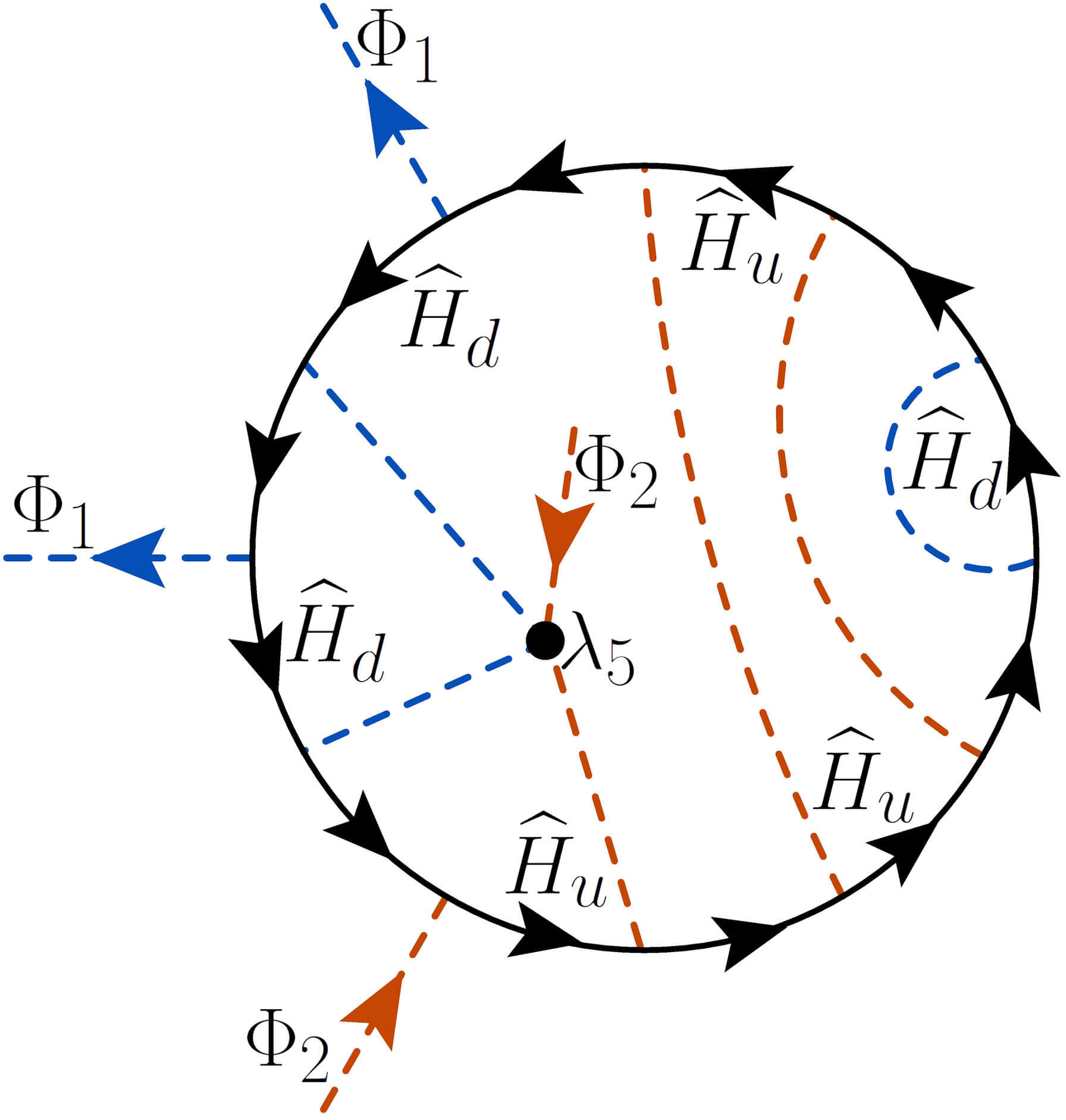}} \,  \resizebox{0.375\linewidth}{!}{ \includegraphics{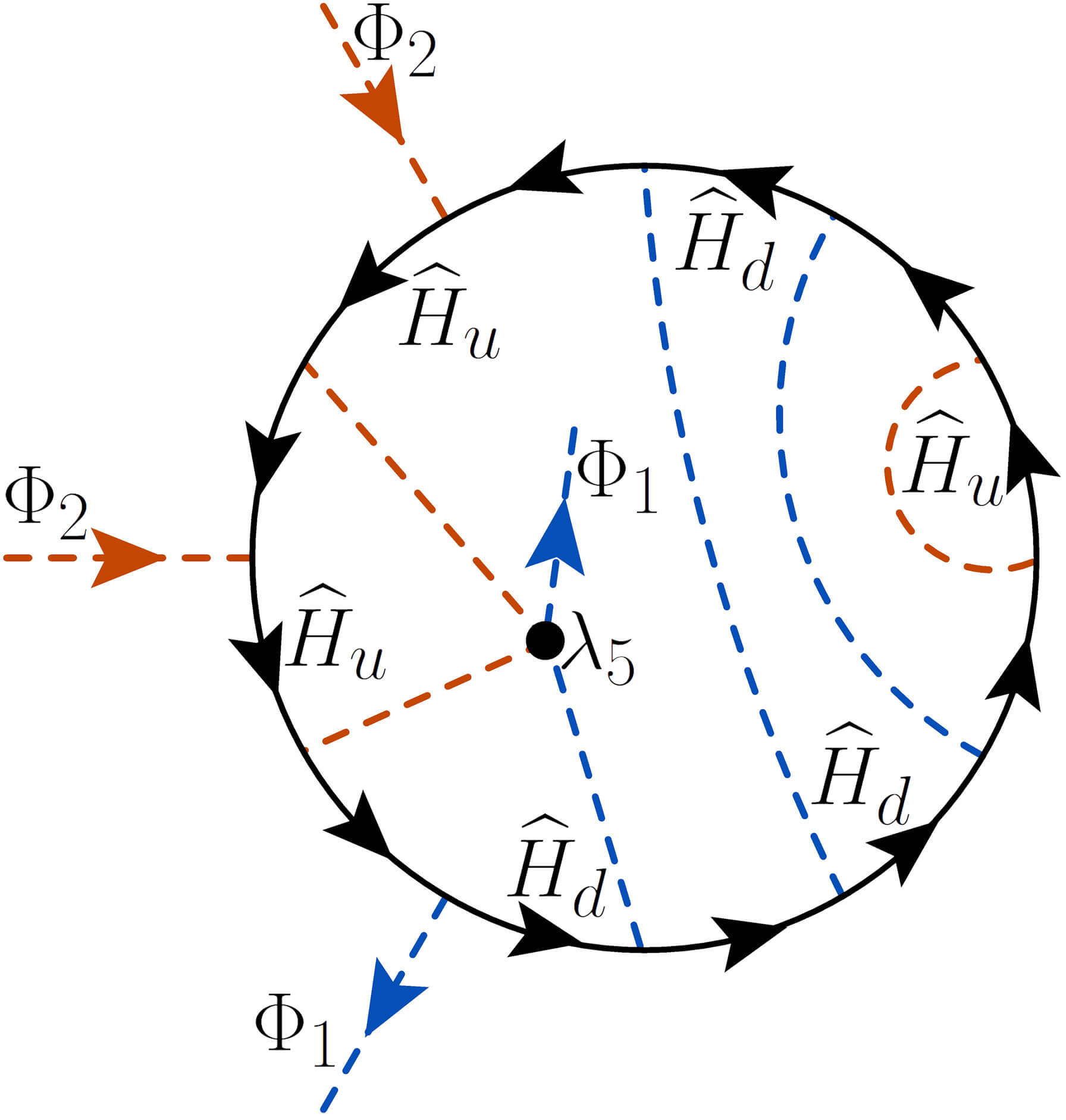}}
\caption{One of the six-loop diagrams proportional to $\mathcal{J}$ (left) and the corresponding diagram proportional to $\mathcal{J}^*$ (right) in the type-II 2HDM, between which the imaginary divergence cancels.}
\label{fig:ex2}
\end{figure}

We show such a pair of diagrams in Fig.~\ref{fig:ex2}. Notice that the cancelation of the imaginary part relies on the fact that the divergent contribution does not depend on the masses of $\Phi_1$ and $\Phi_2$, so that the scalar propagators become indistinguishable in the Feynman integrals. This implies that \emph{finite} imaginary contributions need not cancel, but they do not affect the renormalizability of the real 2HDM. We thus conclude that the type-II 2HDM has no divergent CPV at six loops.

\begin{figure}[h!]
 \resizebox{0.375\linewidth}{!}{ \includegraphics{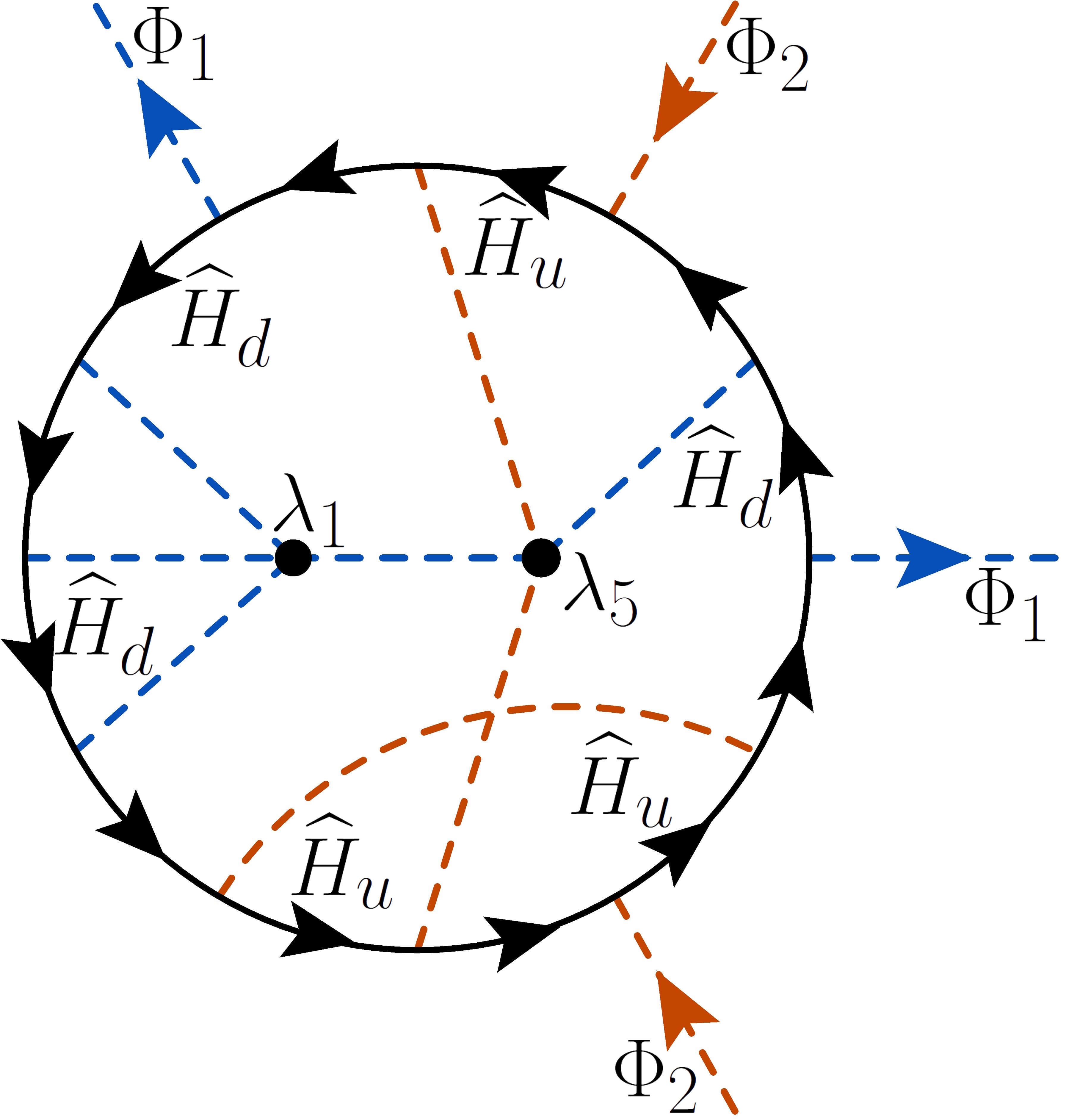}} \,  \resizebox{0.375\linewidth}{!}{ \includegraphics{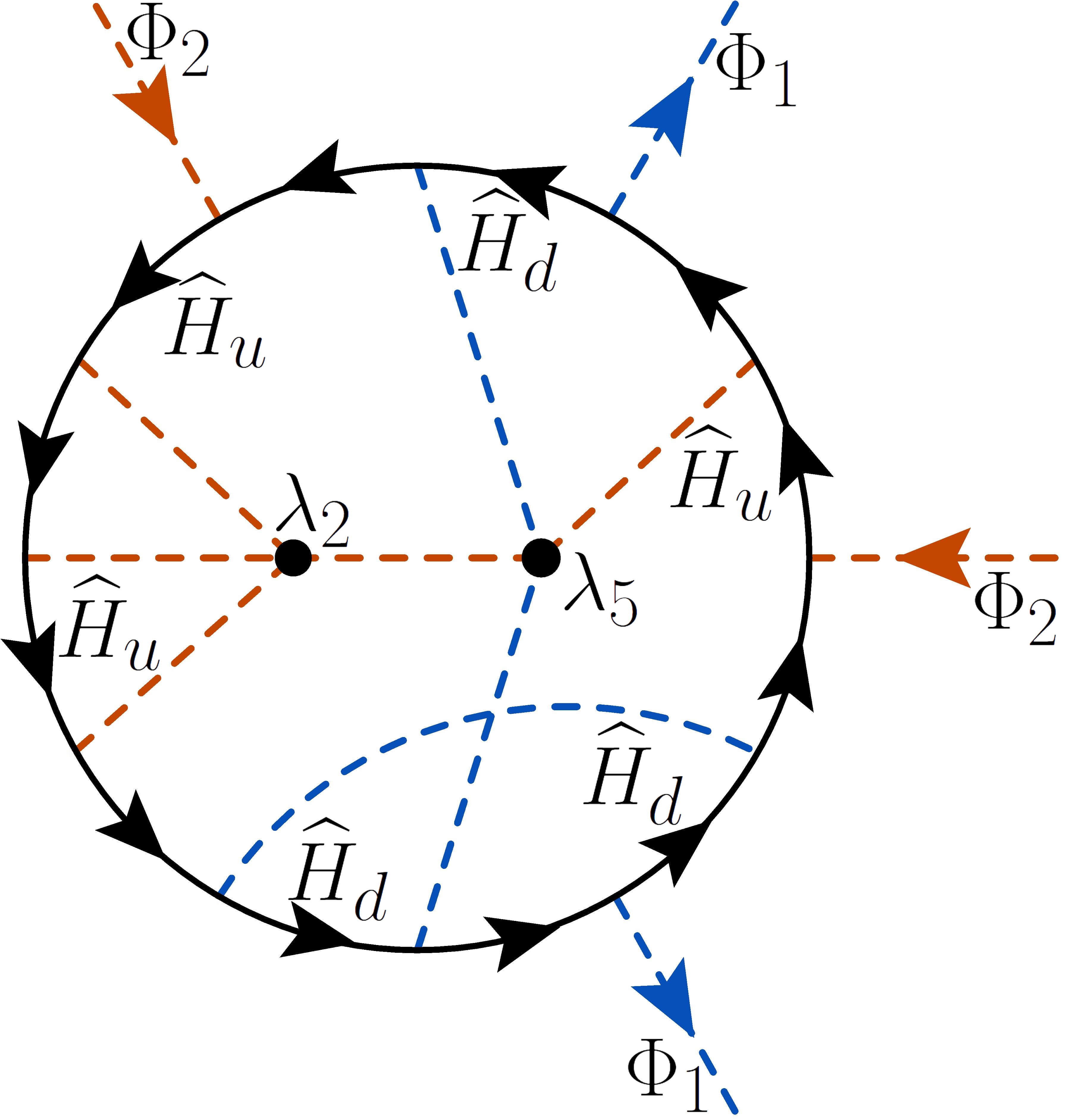}}
\caption{One of the seven-loop diagrams proportional to $\mathcal{J}$ with a $\lambda_1$ insertion (left) and the corresponding diagram proportional to $\mathcal{J}^*$ with a $\lambda_2$ insertion (right) in the type-II 2HDM.}
\label{fig:typeII7loop}
\end{figure}

This generalized CP transformation is an accidental symmetry of the six-loop diagrams because of their limited coupling dependence. When applied to the full scalar potential in Eq.~(\ref{eq:scalarpot}), it interchanges $\lambda_1$ and $\lambda_2$; it is also violated by the unequal hypercharges and Yukawa couplings of $u_R$ and $d_R$. The accidental symmetry is thus not preserved at seven loops, and the reason for canceling the imaginary divergent contribution to $\lambda_5$ is thus removed. Pairs of diagrams proportional to $\mathcal{J}$ and $\mathcal{J}^*$ that do not cancel each other can be constructed by involving one of the following:

\begin{itemize}
\item a hypercharge gauge boson exchange;
\item an additional pair of Yukawa couplings in the quark loop yielding an additional factor of $\widehat H_u$ or $\widehat H_d$ (but not both); 
\item a $\lambda_1$ or $\lambda_2$ vertex (for $\lambda_1 \neq \lambda_2$; see Fig.~\ref{fig:typeII7loop}).
\end{itemize}
We thus expect imaginary divergent contributions to $\lambda_5$ to first appear at the seven-loop level in the type-II 2HDM as well, with coefficients proportional to $\lambda_5 \Im(\mathcal{J})$ times these additional couplings. We have explicitly verified the pairing of diagrams at six loops and the lack thereof at seven loops using QGRAF~\cite{QGRAF} for diagrams without gauge boson insertions. 

%%%%%%%%%%%%%%%%%%%%%%%%%%%%%%%%%%%%%%%%%%%
\section{Conclusion}
\label{sec:conc}
%%%%%%%%%%%%%%%%%%%%%%%%%%%%%%%%%%%%%%%%%%%

In this letter, we provide evidence that the quark sector induces divergent radiative CPV in the scalar potential of the 2HDM. While the necessary ingredients for such radiative CPV are present starting at six loops, we show that approximate symmetries of the contributing diagrams lead to the cancelation of all such divergent CPV at this order. These symmetries are violated by particular types of coupling insertions at seven loops, enabling us to predict that the leading contribution to the renormalization-group (RG) equation~\footnote{The RG equations for the 2HDM up to three loops have been studied in Refs.~\cite{PhysRevD.56.5366,Chowdhury:2015yja,Herren:2017uxn,Bednyakov:2018cmx,Oredsson:2018yho,Oredsson:2019mni,Fontes:2021iue}.} for ${\rm Im}(\lambda_5)$ will have the form:
\begin{widetext}
\begin{align}
 	\frac{d \, {\rm Im}(\lambda_{5})}{d \ln \mu} = 
 	\frac{ \lambda_5 {\rm Im}(\mathcal{J})}{(16 \pi^2)^7} \begin{cases} 
	 \left[ b^{(\lambda_3)} \lambda_3 + b^{(\lambda_4)} \lambda_4 + b^{(g^{\prime})} g^{\prime 2} + b^{(g)} g^2 \right]  &    \, \text{(type I)} \\
	 \\
      \left[ a^{(\lambda)} (\lambda_1 - \lambda_2) 
			+ a^{(g^{\prime})} g^{\prime 2} 
			+ a^{(y)} (y_t^2 - y_b^2 + \ldots) \right]   &  
       \, \text{(type II),} \\
   \end{cases}
\end{align}
\end{widetext}
where $\mu$ is the renormalization scale, $a^{(i)}$ and $b^{(i)}$ are numerical coefficients, $g^{\prime}$ and $g$ are the hypercharge and $SU(2)_L$ gauge couplings, $y_t$ and $y_b$ are the top- and bottom-quark Yukawa couplings, and the ellipses represent smaller terms involving second- and first-generation quark Yukawa couplings.  We emphasize that RG equations for Lagrangian parameters are unaffected by spontaneous symmetry breaking~\cite{Weinberg:1973xwm}, so that working in the broken phase cannot change our conclusion.

The running of $\Im(\lambda_5)$ induced by this seven-loop RG equation is numerically very small. Starting from $\Im(\lambda_5) = 0$ at the Planck scale and running down to the weak scale yields at most $\Im(\lambda_5) \sim 10^{-22}$ for $\mathcal{O}(1)$ real quartic couplings, which is phenomenologically negligible. Likewise, while this CPV contributes to the electric dipole moment of the electron at two loops~\cite{Altmannshofer:2020shb}, its contribution will be many orders of magnitude smaller than the SM contribution.
Nevertheless, our point is a more fundamental one:
the 2HDM with softly broken $Z_2$ symmetry is required to be complex for its own theoretical consistency.

%%%%%%%%%%%%%%%%%%%%%%%%%%%%%%%%%%%%%%%%%%%%
\acknowledgments%\section*{Acknowledgement}
C.H.L.\ thanks the organizers of the 2022 Workshop on Multi-Higgs Models in Lisbon for creating a stimulating environment in which the idea for this project arose. We also thank Pedro Ferreira, Duarte Fontes, Shinya Kanemura, David McKeen, David Morrissey, and Rui Santos for helpful conversations, and Jo\~ao Silva for encouragement.
This work was supported by the Natural Sciences and Engineering Research Council of Canada (NSERC). C.H.L.\ is also supported by TRIUMF which receives federal funding via a contribution agreement with the National Research Council (NRC) of Canada.
%%%%%%%%%%%%%%%%%%%%%%%%%%%%%%%%%%%%%%%%%%%%

\appendix

%%%%%%%%%%%%%%%%%%%%%%%%%%%%%%%%%%%%%%%%%%%%

\bibliography{bib2HCPV}

\end{document}